\documentclass[authorversion, nonacm]{acmart}
\usepackage{wrapfig}
\usepackage{graphicx}
\AtBeginDocument{%
  \providecommand\BibTeX{{%
    \normalfont B\kern-0.5em{\scshape i\kern-0.25em b}\kern-0.8em\TeX}}}

\setcopyright{acmlicensed}
\copyrightyear{2024}
\acmYear{2024}
\acmDOI{XXXXXXX.XXXXXXX}

\acmConference[CHI '24]{Proceedings of the CHI Conference on Human Factors in Computing Systems}{May 11--16, 2024}{Honolulu, HI, USA}
\acmBooktitle{Proceedings of the CHI Conference on Human Factors in Computing Systems (CHI '24), May 11--16, 2024, Honolulu, HI, USA}
\acmISBN{978-1-4503-XXXX-X/18/06}

\begin{document}

\title{Towards Collaborative Family-Centered Design for Online Safety, Privacy and Security}

\author{Mamtaj Akter}
\email{Mamtaj.Akter@vanderbilt.edu}
\orcid{0000-0002-5692-9252}
\affiliation{%
  \institution{Vanderbilt University}
  \city{Nashville}
  \state{Tennessee}
  \postcode{37212}
  \country{USA}
}
\author{Zainab Agha}
\email{Zainab.Agha@vanderbilt.edu}
\orcid{0000-0002-6519-3742}
\affiliation{%
  \institution{Vanderbilt University}
  \city{Nashville}
  \state{Tennessee}
  \postcode{37212}
  \country{USA}
}
\author{Ashwaq Alsoubai}
\email{Ashwaq.Alsoubai@vanderbilt.edu}
\orcid{0000-0003-1569-9662}
\affiliation{%
  \institution{Vanderbilt University}
  \city{Nashville}
  \state{Tennessee}
  \postcode{37212}
  \country{USA}
}
\affiliation{%
  \institution{King AbdulAziz University}
  \city{Jeddah}
  \postcode{96657}
  \country{KSA}
}
\author{Naima Samreen Ali}
\email{Naima.Samreen.Ali@vanderbilt.edu}
\orcid{0009-0006-4446-9689}
\affiliation{%
  \institution{Vanderbilt University}
  \city{Nashville}
  \state{Tennessee}
  \postcode{37212}
  \country{USA}
}

\author{Pamela J. Wisniewski}
\email{Pamela.Wisniewski@vanderbilt.edu}
\orcid{0000-0002-6223-1029}
\affiliation{%
  \institution{Vanderbilt University}
  \city{Nashville}
  \state{Tennessee}
  \postcode{37212}
  \country{USA}
}

\renewcommand{\shortauthors}{Akter et al.}

\begin{abstract}
Traditional online safety technologies often overly restrict teens and invade their privacy, while parents often lack knowledge regarding their digital privacy. As such, prior researchers have called for more collaborative approaches on adolescent online safety and networked privacy. In this paper, we propose family-centered approaches to foster parent-teen collaboration in ensuring their mobile privacy and online safety while respecting individual privacy, to enhance open discussion and teens' self-regulation. However, challenges such as power imbalances and conflicts with family values arise when implementing such approaches, making parent-teen collaboration difficult. Therefore, attending the family-centered design workshop will provide  an invaluable opportunity for us to discuss these challenges and identify best research practices for the future of collaborative online safety and privacy within families.
\end{abstract}






\maketitle

\section{Introduction}

In today's digital world and increased social media usage \cite{Anderson_2023_pewresearch}, teens encounter several online risks ranging from cyberbullying \cite{vogels_2022_pewresearch_cyberbullying}, exposure to explicit content \cite{martinez2023social}, and unwanted sexual solicitations \cite{razi2020let}. Traditional safety solutions to manage teens' online safety focused on restrictive approaches, such as parental controls \cite{modecki2022}. However, teens often found such restrictive approaches to be privacy-invasive \cite{wisniewski2018privacy}, leading to mistrust between parents and teens. Additionally, such restrictive approaches may hinder the development of teens' self-regulation skills needed to navigate online risks. Therefore, several adolescent online safety researchers \cite{wisniewski2013grand, walker2019moving} have called for more teen-centric and collaborative approaches so teens and parents can work together to develop effective online safety strategies. 

Along with these adolescent online risks, the rise in smartphone usage \cite{nw_mobile_2024} and third-party app installations \cite{noauthor_majority_2015} poses mobile privacy and security threats across all age groups. Ironically, the majority of U.S. adults lack sufficient digital privacy knowledge, leading to a lack of control over their data privacy \cite{vogels_americans_2019}. Scholars in networked privacy advocate for collaborative and community-based approaches to empower individuals in managing digital privacy and security \cite{chouhan_co-designing_2019, aljallad_designing_2019, akter_evaluating_2023, akter_CO-oPS_2022, badillo2020towards}. However, implementing such approaches faces challenges due to imbalanced power dynamics and distrust regarding technological expertise \cite{akter_CO-oPS_2022, akter_it_2023}. Our aim in attending this workshop is to explore actionable strategies for integrating family-centered design practices towards these collaborative approaches, to ensure adolescent online safety as well as improved privacy and security within families.

\section{Moving From Restrictive Approaches to Collaborative Family-centered Approaches}
Our research team is presently engaged in multiple funded research projects related to collaborative approaches to online safety and mobile privacy within families \cite{agha2021just, agha2023strike, alsoubai2022mosafely, akter_from_2022, akter_it_2023}. In this section, we outline how our proposed approaches transcend traditional methods centered on restrictive parental control, shifting towards a family-based design that allows both parents and teens equal agency in assisting each other with mobile privacy and online safety. 

\subsection{Towards Ethical Research with Families  to Co-Design Real-Time Online Safety Solutions}

Understanding youths' unsafe online experiences often involves discussing uncomfortable topics. As such, we need to find ethical ways to involve families in research that protect the participants' from harm. To overcome this challenge, researchers are increasingly advocating for meta-level research (or "research on research"), especially when working with vulnerable populations \cite{walker2019moving}. In our work, we applied this methodology with teens and their parents to understand their preferences for participating in research on sensitive topics\cite{badillo2021conducting}. We used two research methodologies (i.e., diary studies and
analyzing social media trace data) as probes to obtain tangible feedback from participants using co-design. We found that both groups wanted to contribute to online safety research on sensitive topics, but had several concerns related to participation. For instance, teens feared getting in trouble with parents or authorities for sharing online risks. Both parents and teens wanted assurances regarding data protection, transparency about the risks, and helpful resources during the research. Overall, teens and parents preferred diary study approaches over the collection of social media trace data, as it granted them more control over what they shared with researchers.

As such, we conducted a repeated measures web-based diary study with parent-teen dyads to better understand the influences between parental mediation and adolescent online risks \cite{agha2021just}. While prior literature largely focused on how parents influence teens' online safety \cite{sasson2017role, shin2012tweens}, we took a holistic approach to understand family dynamics on how parents and teens impact \textit{each other} regarding online safety on a weekly basis. We found that parental mediation and teen online risk exposure were most significantly correlated in the same week, suggesting parenting occurred ‘just-in-time,’ for teens' online risks. Such "real-time" interventions can serve as teachable moments for teens in-the-moment they face a risk. Our findings also emphasize the need for more collaborative approaches within families that can help manage their online safety collectively. Our team has focused on several efforts for co-designing and evaluating real-time interventions (e.g., "nudges" \cite{thaler2009nudge, agha2022case}) with teens User Experience (UX) bootcamps \cite{agha2023strike, agha2022case}, and Youth Advisory Boards (YAB) \cite{ali2024case}. Most teens developed features that were built upon the idea that social media platforms provide accurate and automated risk detection in real-time. Therefore, our team has led two other branches of this work summarized below, including a) collaborative approaches for managing parents' and teens' online privacy, and b) automated risk detection that is a precursor for effective real-time interventions and family-centered technologies. 

\subsection{Joint Family Oversight Approaches to Mobile Online Safety, Privacy and Security}

The widespread use of smartphones and third-party apps \cite{nw_mobile_2024, noauthor_majority_2015}, combined with a significant lack of data privacy knowledge among most US adults \cite{vogels_americans_2019}, underscores the necessity for collaborative approaches \cite{chouhan_co-designing_2019, kropczynski_examining_2021, akter_evaluating_2023, akter_CO-oPS_2022} to ensure mobile online safety and privacy for people of all ages, including adolescents and adults. Our team has recently proposed a joint family oversight approach to mobile online safety and privacy, enabling parents and teens to collaborate in ensuring safe smartphone usage \cite{akter_from_2022}. We developed the "CO-oPS" app based on a community oversight model for privacy and security, allowing family members to review installed apps and privacy permissions, hide certain apps, and provide feedback \cite{chouhan_co-designing_2019, akter_CO-oPS_2022,  akter_evaluating_2023}. Through a lab-based study with 19 parent-teen dyads \cite{akter_from_2022}, we explored their current approaches to managing mobile online safety and app privacy, and perceptions regarding the applicability of CO-oPS app within their family context. Our findings revealed a lack of consideration among parents and teens regarding mobile online safety and privacy when installing new apps or granting permissions, with parents manually monitoring their teens' app usage, and teens showing minimal interest in ensuring their parents' mobile privacy and security. While both parents and teens appreciated the awareness of one another's app usage and permission settings provided by the CO-oPS app, they were less enthusiastic about its privacy feature that allowed them to hide their apps from one another, citing concerns about its impact on transparency-based relationships. Power imbalances between parents and teens also emerged as a challenge, with parents more open to joint co-monitoring and teens hesitant to monitor their parents' mobile apps and privacy practices. To address these challenges, we suggested design recommendations such as nudges to improve communication about mobile privacy and security, pro-tips, and expert advice to enhance parents' privacy knowledge, and incentive mechanisms to encourage teens' participation in family oversight. Overall, our findings show that increased transparency could enhance discussion and mutual learning, ultimately improving family online safety and digital privacy. However, achieving these benefits relies on shared responsibility between family members for each other's online safety and privacy, a paradigm shift from individualistic approach or traditional parental control. Furthermore, the insights gained from utilizing CO-oPS within familial settings \cite{akter_from_2022} can provide benefits not only to developers creating similar collaborative family online safety tools but also to designers devising tactics to aid users in protecting their digital privacy. This expands beyond the realm of mobile online safety to include broader realms of digital privacy and security, such as smart home devices \cite{alghamdi_codesigning_2023, alghamdi_misu_2022, emami_influence_2018}, social media \cite{bhagavatula_adulthood_2022, such_privacy_2016, vishwamitra_towards_2017, ulusoy_panola_2021}, websites \cite{mcdonald_citizens_2021, mcdonald_building_2021}, and other platforms where adolescents and their families disclose personal information to third parties.

\subsection{MOSafely is that Sus? A Collaborative Approach to Evaluate Automatic Risk Detection Algorithms}
The integration of Artificial Intelligence (AI) highlights significant advancements in developing technologies for detecting adolescent online risks ~\cite{wronska2020education}. A common limitation is that many tools designed for detecting risks among youth are not accessible for evaluation by the public, often resulting in a high rate of false positives~\cite{razi2021human, alsoubai2024profiling,samek2019explainable}. Thus, the involvement of key stakeholders is crucial to validating the accuracy of detection models and enhancing these models through their feedback.
Compounding this issue is the observed discrepancy between how parents and youths perceive online risks~\cite{wisniewski2017parents}. Research has shown a communication gap between these groups, leading to a misunderstanding about the nature and frequency of these risks ~\cite{wisniewski2017parents,symons2020parents,agha2021just}. Parental control software has largely been unsuccessful in bridging this gap~\cite{akter_from_2022}. This highlights the need for more collaborative and youth-centered approaches that improve parent-teen communication and encourage self-regulation by teens regarding their online risks.

In response to these intertwined challenges, the MOSafely's Is that Sus? dashboard emerges as a pivotal solution~\cite{alsoubai2022mosafely}. This dashboard is tailored to incorporate pre-trained machine-learning algorithms capable of identifying various online risks, such as sexual solicitation, cyberbullying, and issues related to mental health~\cite{alsoubai2022friends, razi2023sliding, ali2022understanding}. It allows youth to upload their social media content, including both public and private interactions, for AI-based risk evaluation. Youth can then examine the risks flagged by the AI within their private conversations to assess and reflect on the validity of these automatically identified risks. Then, youth have the option to share selected AI-identified risks along with any misclassification with their parents, who can then offer their own insights on the algorithm's outcomes and the perceived risks. After this phase, researchers will conduct interviews with both youth and their parents to delve into their individual viewpoints and work towards resolving any discrepancies. While this study is still in progress,  by adopting a family-centric approach, we seek to address the need for stakeholder involvement in the evaluation of risk detection algorithms and bridging the communication gap between youths and their parents. 

\section{Conclusion}

By attending this workshop, our objective is to actively participate in discussions with fellow researchers, delving into best practices for implementing family-based collaborative approaches outlined in our position paper. Our aspirations for attending include: 1) Establishing a supportive network of researchers to exchange experiences and foster collaboration; 2) Acquiring actionable insights to integrate innovative family-centered design practices into our research on mobile privacy and adolescent online safety, thereby enhancing its impact and relevance.
\begin{acks}
This research is partially supported by the U.S. National Science Foundation under grants \#CNS-2326901, \#IIP-2329976, \#IIS-2333207 and by the William T. Grant Foundation grant \#187941. Any opinions, findings, and conclusions or recommendations expressed in this material are those of the authors and do not necessarily reflect the views of the research sponsors.
\end{acks}

\section*{Author Bios}

\noindent
\textbf{Mamtaj Akter} is a Ph.D. candidate in Computer Science at Vanderbilt University. Her research explores how community oversight plays a role on improving individuals’ mobile online safety, digital privacy and
security decisions. \\

\noindent
\textbf{Zainab Agha} is a Ph.D. candidate in the Department of
Computer Science at Vanderbilt University. Her research takes a teen-centric approach to online safety, focusing on parent-teen collaboration in co-designing real-time online safety interventions. \\

\noindent
\textbf{Ashwaq Alsoubai} is an Assistant Professor at the Department of
Computer Science at King Abdul Aziz University. Her research is to improve adolescent online safety utilizing human-centered machine learning approaches. \\

\noindent
\textbf{Naima Samreeen Ali} is a second year Ph.D. student in the Department of
Computer Science at Vanderbilt University. Her research focuses on effectively involving teens in online safety research to inform future interventions aimed at keeping teens safe online. \\

\noindent 
\textbf{Pamela Wisniewski} is an Associate Professor and Flowers Family Chancellor Faculty Fellow of Computer Science at Vanderbilt University. Her current research interests include Human-computer Interaction, Social Computing, and Adolescent Online Safety. Dr. Wisniewski received her Ph.D. degree in Computer and Information Systems from the University of North Carolina at Charlotte. She is an ACM Senior Member. 

\bibliographystyle{ACM-Reference-Format}
\bibliography{00_MAIN}


\end{document}